\documentclass{aa}  

\usepackage{natbib}
\bibpunct{(}{)}{;}{a}{}{,} 

\usepackage{graphicx}
\usepackage{txfonts}

\usepackage[normalem]{ulem}
\newcommand{\rv}[1]{#1}
\newcommand{\rvd}[1]{}

\begin{document} 
\title{A helical jet model for OJ287}
\author{M. Valtonen\inst{1,2}
  \and
  P. Pihajoki\inst{3}}

\institute{
FINCA, University of Turku, 21500 Piikki\"{o}, Finland\\
\and 
Department of Computer Science, Mathematics and Physics, The University of the West Indies, Barbados \\
\and
Department of Physics and Astronomy, University of Turku, 21500 Piikki\"{o}, Finland \\
\email{popiha@utu.fi}
}

\date{Received January 00, 2013; accepted January 00, 2013}

\abstract
{\object{OJ287} is a quasar with a quasi-periodic optical light curve, with the
periodicity observed for over 120 years. This has lead to a binary black
hole model as a common explanation of the quasar. The radio jet of \object{OJ287}
has been observed for a shorter time of about 30 years. It has a
complicated structure that varies dramatically in a few years time
scale.} 
{Here we propose that this
structure arises from a helical jet being observed from a small and
varying viewing angle.
The viewing angle variation is taken to be in
tune with the binary orbital motion.}
{We calculate the effect of the secondary black hole on the inner edge
of the accretion disk of the primary using particle simulations. We
presume that the axis of the helix is perpendicular to the disk.
We then follow the jet motion on its helical path and project the jet to
the sky plane. This projection is compared with observations both at mm
waves and cm waves.
}
{We find that this model reproduces the observations well if the
changes in the axis of the conical helix propagate outwards with a
relativistic speed of about $0.85 c$. In particular, this model explains
at the same time the long-term optical brightness variations as varying
Doppler beaming in a component close to the core, i.e. at parsec scale
in real linear distance, while the mm and cm radio jet observations are
explained as being due to jet wobble at much larger (100 parsec scale)
distances from the core.}
{}


\keywords{BL Lacertae objects: individual (OJ287) – quasars: individual (OJ287)}

\maketitle

\section{Introduction}

Optical brightness measurements of the quasar \object{OJ287} have shown cyclical
behavior with two well-defined cycles: a 60 year cycle and a 12 year cycle
\citep{val2006}. In 1995 a model was constructed with gave a good
account of the 12 year cycle, occurring in two major bursts separated by 1 – 2
years 
\citep{leh1996,sun1997}. The model predicted the
optical brightness level for every two-week interval from 1996 until 2030, and
also explained the brightness of \object{OJ287} at similar steps from 1900 up to 1996.
Even though the number of historical data points has increased by an order of
magnitude since 1995, the calculated two-week average points still give a fair
explanation of the past behavior 
\citep{hud2013}. As to the future
predictions, the two-week points have been correct from 1996 to 2010, with
surprisingly small errors.
Other suggested models,
producing a quasi-periodic 12 yr cycle, can be excluded at 5 sigma significance
level at present 
\citep{val2011}.
When time goes on, and if \object{OJ287} continues to produce the
expected optical magnitudes at each two-week step, then this confidence level
will go up.

A new challenge has arisen with the complex behavior of the resolved radio jet.
There it is possible to identify the same 12 yr cycle in the position angle of
the radio jet in the sky in the cm wave radio maps 
\citep{tat2004}.
Moreover, the 12 year cycle is clearly modulated by a longer time scale
variation 
\citep{val2012b} even though it is too soon to claim that the
60 year cycle is also detected since the radio jet observations cover only
about 30 years, in contrast to the 120 year optical light curve. A new twist in
the story is the radio jet measurements at mm waves: they show variations of
much larger amplitude 
\citep{agu2012,tat2013}.
The first radio maps
at these frequencies go back only less than twenty years, and thus it would not
be useful to model these data alone. However, a combination of optical
variability data together with the jet variations at these two frequency ranges
provides enough challenging data to attempt a model.

\citet{val2012b} looked at the variations of the rotation axis of the
inner accretion disk in the well-defined binary model which gives detailed
account of the optical variations 
\citep{val2008}. 
They found that indeed the disk shows a wobble in 12 year time scale that could be associated
with the jet variations at cm wavelengths. They also found that a small timing
delay of about 16 years is necessary between the disk wobble and the jet
changes, in order to give the best account of the data. The model produces two
cycles, a short one of about 11 year period, and a long cycle of about 120
years. The latter is understood as the Kozai cycle in the inner accretion disk,
inside the pericenter of the binary orbit 
\citep{inn1997}. Subsequently,
\citet{val2012c}
investigated the possibility of including the mm wave
observations in the same model. They found that the jet cannot be straight in
order that the different orientation angles can arise at the same time.

This leads us to a jet model which has been found useful in \object{S5 0836+710} and in
other jet sources 
\citep{per2012,per2013}: a helical jet. This is a way,
even if not a unique way, of producing different viewing angles at different
distances from the origin of the jet. When we add to this the disk wobble which
arises from the model based exclusively on optical data, we get a new attempt
to model the complex situation in \object{OJ287}.

\section{The inner disk}

Our basic model is a binary black hole system where a small companion black
hole perturbs periodically the disk of the much larger primary black hole
\citep{val2010}. The parameters are very tightly constrained by the
optical data which leaves practically no free parameters in the model. The only
exception is the orbital inclination of the secondary orbit relative to the
disk which has been unknown until recently. Its determination requires the
identification of the spectral lines of the secondary and the
measurement of the relative radial velocity. From the recent detection of such a line
(T. Pursimo, private communication) we determine the inclination to be close to
50 degrees. Previously the inclination of 90 degrees was used for simplicity;
here we calculate both the 50 and 90 degree cases. 

\begin{figure}
\centering
\includegraphics[angle=-90,width=\hsize]{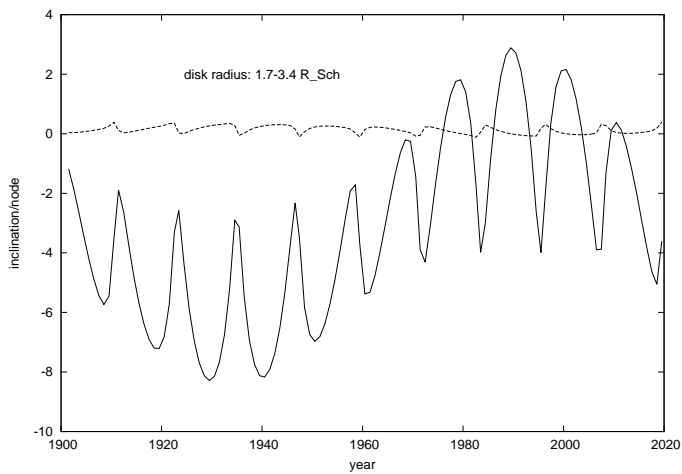}
\caption{\label{fig:fig1}
The variation of the inclination with respect to the
initial value of 50 degrees (solid line) of the inner accretion disk, and the
variation of the node with respect to initial value of zero (dashed line).
}
\end{figure}

We place a ring of particles around the primary, and starting from year 1856,
follow the evolution of each disk particle. 
\rv{The calculations are carried out as in \citet{val2012b}:
In the present simulation the number of disc particles is
1500. 
They are placed in circular orbits between 1.7 and 14 Schwarzschild
radii of the primary.
For every particle, and for every time step, we
calculate the orbital elements of the orbits with respect to
the primary. The elements are averaged per calendar year.
As there are many integration steps per year, typically each
annual mean is based on the average of $10^5$
values. By varying the number of particles it
was found that the mean values generated in this way are
very robust.}
\rvd{At regular intervals, once per
year, we calculate the mean values of the orbital elements. Each particle
enters the average several times a year, at every integration step 
\mbox{\citep[see][]{val2012b}}. 
The parameter values obtained this way are extremely stable.}
Figure~\ref{fig:fig1} shows the time evolution of two of the elements,
inclination $i$ and ascending node $\Omega$. 

\begin{figure}
\centering
\includegraphics[angle=-90,width=\hsize]{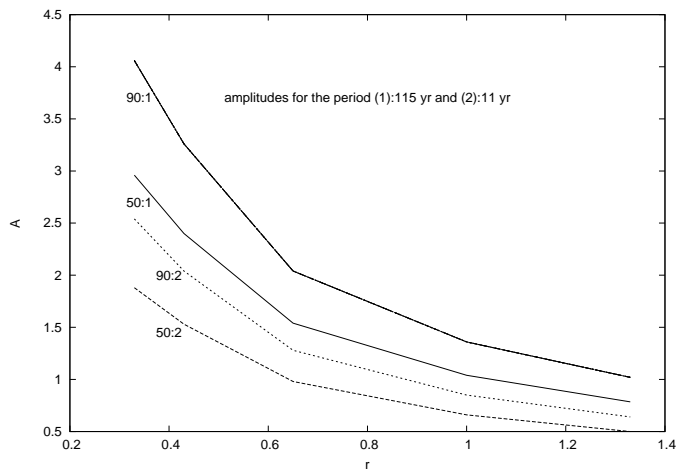}
\caption{\label{fig:fig2}
The amplitudes $A_1$ and $A_2$ (in degrees) of
the two periodic components of the inclination variation as a function of the
distance from the centre $r$ in units of 10 Schwarzschild radii of the primary. 
The first number in the label gives the initial inclination in degrees,
the second number the subscript of the amplitude coefficient.
}
\end{figure}

As in 
\citet{val2006}, the variation of $\Omega$ is negligible in
comparison with $i$ (Valtonen and Wiik 2012 had interchanged $i$ and $\Omega$
by mistake; this however had no consequences in that paper apart from wrong
labels. Correcting this error, the results agree with the present work.). The
evolution of the element $i$ may be well represented by a doubly periodic
function
\begin{equation}
i - i_0  = -A_1\sin\left[2\pi(t-t_1)/P_1\right] - A_2\sin\left[2\pi(t-t_2)/P_2\right] - C.
\end{equation}
The amplitudes $A_1$ and $A_2$ depend on the initial inclination $i_0$ as well as on
the distance $r$ of the ring from the primary black hole. The values of the
coefficients are shown in Fig.~\ref{fig:fig2}.  

We are particularly interested in the inner edge of the disk which presumably
connects to the jet which is launched perpendicular to the disk. Therefore we
will use the values of $A_1$ and $A_2$ which are appropriate for the inner disk, 
with $r$ between $1.7$ and $3.4$ Schwarzschild radii. The smaller value
corresponds to the last stable orbit in the accretion disk.
We also consider the case of $i_0=50$ degrees only, with $A_1=3.5$. We notice from
Fig.~\ref{fig:fig2} that the amplitudes are somewhat greater for
$i_0=90$ degrees than for $i_0=50$ degrees. The
calculation below would lead to small changes in the jet parameters if we used
$i_0=90$ degrees instead of $i_0=50$ degrees. The two periods $P_1=116.6\ \mathrm{yr}$ and
$P_2=11.0\ \mathrm{yr}$ are insensitive to $i_0$.

One of the differences between a gaseous disk and the simulated particle disk
is the need to transport information between particles in order to mimic gas
behavior. We do this by simply averaging the results over a period longer than
one year in order to allow time for dissipative processes. The one year
averaging was arbitrarily chosen. Considering that the period rotation of the
innermost disk is about 0.5 yr, and that the viscous time scale of the inner
disk is about an order of magnitude greater than this, the appropriate
averaging time scale is in the range of 5 to 10 years 
\citep{kro2005}. 
We use 10 yr averaging 
\citep[as in ][]{val2012b}
which lowers the coefficient
$A_2$ from $2.24$ to $0.75$. The values for $t_1=1903.9$ and $t_2=1893.5$ are
insensitive to the inclination $i_0$ and to the averaging time scale while the
value of $C=3.13$ is of no importance to what follows. 

\section{The helix model}

We assume that a helical jet emanates from the center of the accretion disk,
perpendicular to the disk. This should be true independent of the direction of
the spin of the primary black hole 
\citep{pal2010,mck2013}. The jet flow is
assumed to follow a helical path, due to helical Kelvin-Helmholtz instability
\citep{har2000,har2011,mig2010,miz2012}. 
The parameters of the helix, the half-opening angle of the
helix cone $\theta$ and the wavelength $\lambda$  of the helix are to be
determined by fitting into the observational data. Also the viewing angle $\phi$
of the jet at a given time and at a given distance from the origin is a free
parameter; however, the evolution of the viewing angle follows from the wobble of
the disk, and contains only one free parameter, the time delay $d$ between the
changes in the disk and the corresponding changes at the emitting region
\citep{val2012c}.

\begin{figure}
\centering
\includegraphics[angle=-90,width=\hsize]{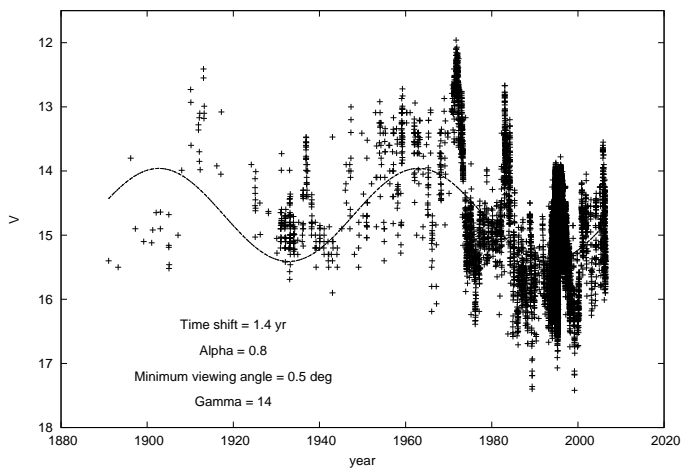}
\caption{\label{fig:fig3}
Optical V-magnitude of \object{OJ287} as a function of time. The line is a fit of the
long-period component of the model. The model parameters are given in the
figure.
}
\end{figure}

We will first consider the region of optical emission in the jet. After
experimenting with different parameter values, we find that the overall
brightness evolution is well modeled if $\theta=4$ degrees, and the minimum
viewing angle of the cone axis is  $\phi_\mathrm{min}=1.8$ degrees. 
\rv{
As the first step to obtain these values, the best sinusoidal long-period fit
to the optical magnitude data was determined. This fit is shown in
Figure~\ref{fig:fig3}. Then the angular parameters of the helical model were
varied at intervals of 0.1 degrees, and for each model a light curve was
produced, and for each theoretical light curve a similar sinusoidal fit was
performed. The best match with the observations was obtained with the above
mentioned values. Their uncertainty may be estimated as $\pm0.1$ degrees.
}
\rvd{
Figure~\ref{fig:fig3} shows the
fit of the long-period cycle to the observations.  
}

\begin{figure}
\centering
\includegraphics[angle=-90,width=\hsize]{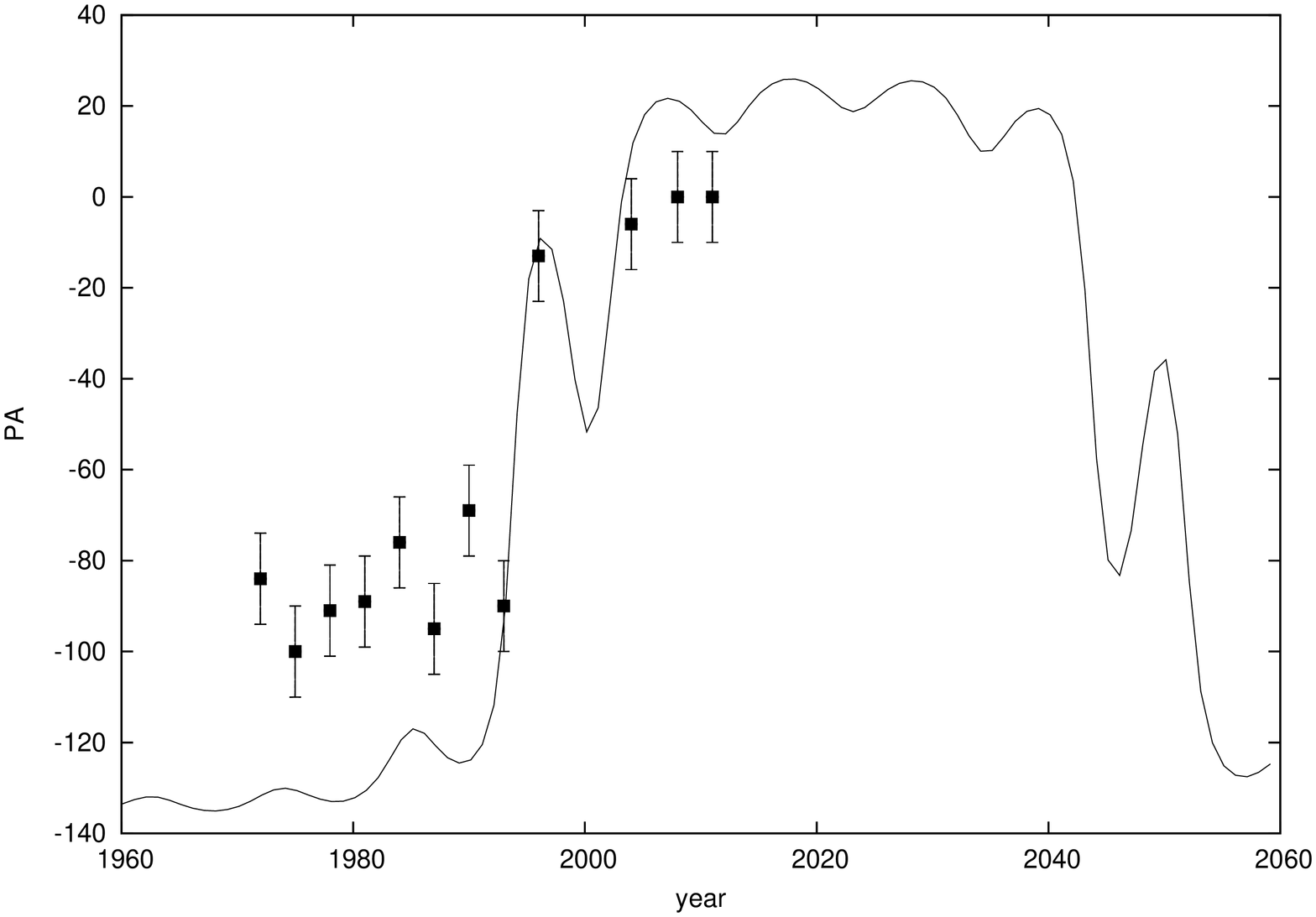}
\caption{\label{fig:fig4}
The evolution of optical polarization in \object{OJ287}. 
Data points are based on
\rv{\citet{val2012b}}, 
while the curve corresponds to a model where the electric vector of the
radiating electrons lies perpendicular to the jet direction in the optical
emission region.
\rvd{
The larger amplitude of variation in the model as compared
with data may signify the existence of a second steady component.}
}
\end{figure}

The time delay $d=1.4\ \mathrm{yr}$. It could also be $58.3+1.4\ \mathrm{yr}$, i.e. with a shift of
one half of the cycle period. However, the distance of the optically emitting
region is more likely to be around $1.4\times\Gamma$ lyr than
$60\times\Gamma$ lyr from the center 
\rv{
\citep{agu2012b,kus2013}.
} 
Moreover, the rather abrupt jump of the optical polarization by $\sim 90$
degrees around 1995 occurs only in the model with the shorter time delay (Fig.~\ref{fig:fig4}). 
\rv{
The theoretical values in Figure~\ref{fig:fig4} were calculated assuming that
the electric vector of the radiating electrons is always perpendicular to the
jet and the assumed prevailing parallel magnetic field. The jump in calculated
values is noticeably larger than observed.  However, as discussed in
\citet{val2012b}, there is a possibility that the electric vector is not
strictly perpendicular to the jet direction, but at a small
angle to it. This is expected as the optical emission is likely to originate
from a compression in the jet, which in turn strenghtens the
perpendicular component
of the magnetic field, and causes the magnetic field direction to bend at the
compression. In the case of \object{OJ287}, the magnetic field along the jet has
been observed to be vary all the way from parallel to perpendicular
\citep{gab2001}.
If we do assume an angle offset of about 15 degrees, the calculated
values match the observations.
}

In this model the magnetic field of the optical emission region lies
more or less parallel to the jet, as was concluded also in previous work 
\citep{val2012b}.
The required Lorentz $\Gamma=14$, in agreement with other determinations (Jorstad
et al. 2005, $\Gamma=16.5\pm4$; Hovatta et al. 2009, $\Gamma\sim15.4$;
Savolainen et al. 2010, $\Gamma\sim9.3$; Agudo et al. 2012, $\Gamma=14\pm5$).
Also the viewing angles reported from observations fall within the range of the
model 
(\citealt{jor2005}, $\phi=3.2\pm0.9$ degrees; 
\citealt{hov2009}, $\phi\sim3.3$ degrees; 
\citealt{sav2010}, $\phi\sim1.9$ degrees; 
\citealt{agu2012}, $\phi=2\pm1.3$ degrees). The spectral index is $-0.8$, which takes into
consideration of the internal extinction in \object{OJ287} host galaxy 
\citep{val2012}.

\begin{figure}
\centering
\includegraphics[angle=-90,width=\hsize]{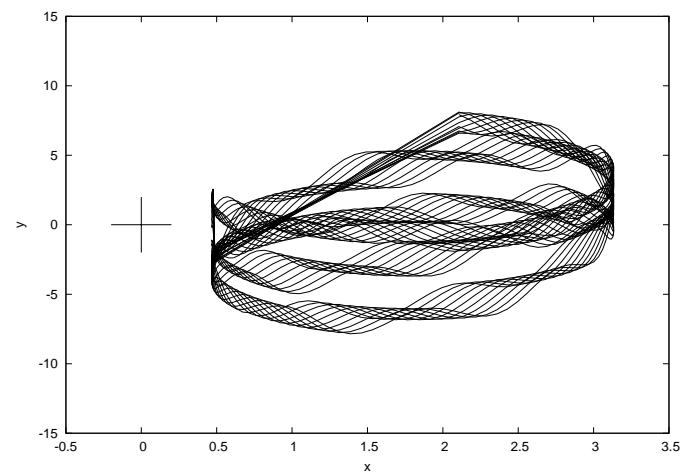}
\caption{\label{fig:fig5}
The path of the jet in the sky plane. The
different lines refer to slightly different time delays. The x-axis corresponds
to $\Omega$ and the y-axis to $i$, both given in degrees. 
The centre of the helix is at (1.8,0), and
there is a foreshortening factor 3.5 which flattens the circular jet profile
along the x-axis to an ellipse. The jet  is taken to move to the positive
x-direction. The cross marks the origin of the jet.
}
\end{figure}

Since the properties of the helix cone are now fully determined by the optical
data, we may apply the model to radio data.
Figure~\ref{fig:fig5} plots the rotation of jet direction in the sky.  

\begin{figure}
\centering
\includegraphics[angle=-90,width=\hsize]{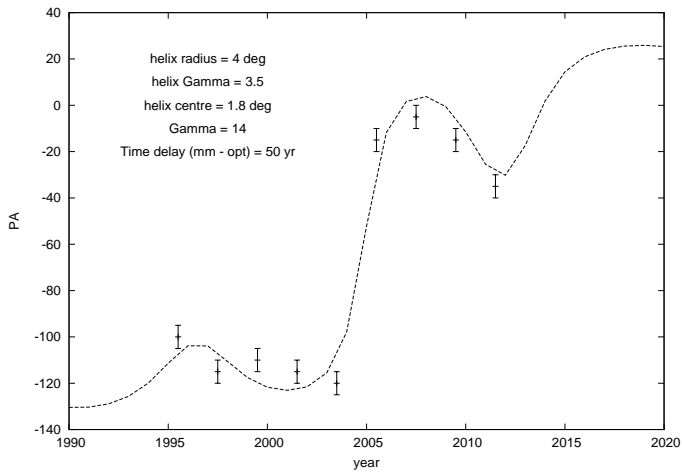}
\caption{\label{fig:fig6}
The position angle of the radio jet at mm waves, shown as points with error
bars, as compared with the model. The model parameters are given in the figure.
}
\end{figure}

The x-axis represents $\Omega$ while the y-axis represents $i$. 
The position angle of the jet is calculated as $\arctan(y/x)$.
These position angles are plotted in Fig.~\ref{fig:fig6} to show the fit to the mm wave
data points 
\citep{agu2012}.  

\begin{figure}
\centering
\includegraphics[angle=-90,width=\hsize]{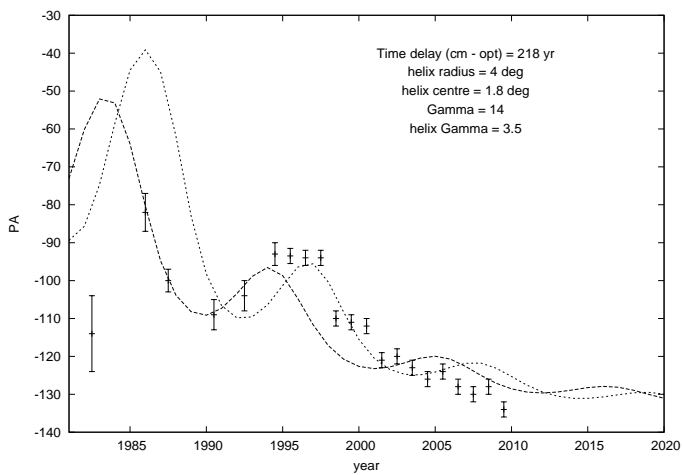}
\caption{\label{fig:fig7}
Comparison of the observed position angle of the cm wave radio jet,
points with error bars, with the model at two different time delays (the two
curves). The model parameters are described in the figure. 
}
\end{figure}

The time delay is 50 yr with respect to optical. It could also be
$50+116.6\ \mathrm{yr}$, but we choose the smallest possible value. For the cm wave
data 
\citep{val2012b}, the fit is shown in Fig.~\ref{fig:fig7}. 

The two curves in the figure present a 3 yr difference in the timing. Since the
resolution of the observations has improved during the 30 year period, it is
expected that the early part and the latter part of the observations would fit a
slightly different model. Overall, the time delay with respect to optical data
is $\sim218\ \mathrm{yr}$. This is a much greater value than was considered by 
\citet{val2012b}. The value could be even greater, $218+116.6\ \mathrm{yr}$, but we choose again
the smallest possible delay.

\begin{figure}
\centering
\includegraphics[angle=0,width=\hsize]{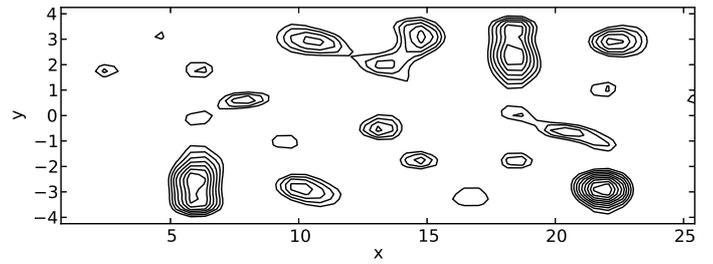}
\caption{\label{fig:fig8}
A jet moving \rv{on the sky plane} in the $x$-direction in the model. 
\rvd{Plots are made for several time delays.} 
\rv{Discrete points in the model jet spanning several helix wavelengths were
calculated and projected to the sky plane.  The sky plane was binned in 15 bins
in $x$ and $y$-direction and smoothed by a factor 3 using cubic splines.
Density contours were then calculated with 10 levels of contours.}
The right hand edge represents the cm wave scale while the left hand
edge represents the mm wave scale. 
}
\end{figure}

\rv{In Figure~\ref{fig:fig5}, the jet direction is taken at the point in the helix which corresponds to
$x=0.5$. This scale is subsequently associated with the mm wave jet.
It then leads to a definite physical scale (see the discussion below).
}
\rv{
In Figure~\ref{fig:fig5}  we trace a point in the jet in the sky plane as a function of time.
Ten different (closely spaced) values of time delay are used. No forward motion
of this point is included, but if you imagine such a motion towards the right
in the figure, then one may guess that the outward moving structure would look
like multiple streams. In Figure~\ref{fig:fig8} the outward motion is included, and the
positions projected in the sky are averaged in the form of a contour map of
the density of points.}

\rvd{
The appearance of the jet in the sky may be inferred from Fig.~\ref{fig:fig5}. You may
imagine the pattern moving towards the positive x-direction. It creates a
structure which looks like two jet streams starting close to the centre, at two
ends of a “bar” perpendicular to the jet (Fig.~\ref{fig:fig8}).  
}
This is also the observed pattern at high resolution 
\citep{tat2013}.  

\rv{
The only unsolved helix parameter remaining is the helix wavelength. 
The cm and mm emitting regions are $218-50\ \mathrm{yr} = 168\
\mathrm{yr}$ apart in the model, which is $168/116.6\sim1.4$ cycles.
We later use this value to find the wavelength, 
after the positions of the mm and cm regions are estimated in the following.}

\section{Jet properties}

Using the typical viewing angle $\phi=2$ degrees, the deprojection factor is
$\sim30$. The projected linear scale of the jet in the sky is $\sim4.6$
parsec per milliarcsec. The mm wave jet is observed with resolution which is
$\sim0.2$ milliarcsec, while the cm wave jet has the resolution of $\sim1$
milliarcsec. Thus the apparent speed at which the changes at the foot of the
jet proceed outwards is $\sim 28\ \mathrm{pc}/50\ \mathrm{yr} \sim 1.7 c$. The
value calculated for the cm wave jet, $\sim 130\ \mathrm{pc}/218\ \mathrm{yr}
\sim 2 c$  is similar.  Thus the kink also moves relativistically.

We may estimate the kink speed as follows. The time it takes the VLBI knots to
move through the mm wave jet is $\Delta t_\mathrm{knot}\sim 1\ \mathrm{yr}$. This may be
compared with the kink travel time $\Delta t_\mathrm{kink}\sim 50\ \mathrm{yr}$ for the same
distance. These quantities are functions of the relative velocities
$\beta_\mathrm{knot}$ and $\beta_\mathrm{kink}$, respectively, and of the viewing angle
$\phi$. The two time intervals are related by
\begin{equation}
\frac{\Delta t_\mathrm{kink}}{\Delta t_\mathrm{knot}} = 
\frac{\beta_\mathrm{knot}(1-\beta_\mathrm{kink}\cos\phi)}{\beta_\mathrm{kink}(1-\beta_\mathrm{knot}\cos\phi)}.
\end{equation}
Putting $\phi=2$ degrees, $\beta_\mathrm{knot}=0.9974$ (corresponding to $\Gamma=14$), and
\begin{equation}
\frac{\Delta t_\mathrm{kink}}{\Delta t_\mathrm{knot}} \sim 50,
\end{equation}
we get
\begin{equation}
\beta_\mathrm{kink}=0.85
\end{equation}
corresponding to $\Gamma\sim3.5$. Thus the true time delay is $\sim175$ yr between
optical and mm waves. The corresponding true delay between mm and cm waves is
$\sim 630\ \mathrm{yr}$. Since the two emitting regions are \rv{1.4} cycles apart in the
model, the true wavelength of the helical wave is about \rv{70 pc}. Considering the
projection factors, this corresponds to about \rv{0.5} milliarcsec separation in the
sky. Sinusoidal features with this wavelength may be recognized in the maps of
\citet{tat2013}. 

\section{Discussion}

\citet{per2012}
discuss fitting a helical jet model to \object{S5 0836+710} which
has wiggly jet. Even though in their case the source of the wiggles is not
known, they discuss parameters that could lead to the observed pattern. These
parameters are not very different from the parameters that we found for \object{OJ287}.
The big advantage in the case of \object{OJ287} is the well specified origin of the jet
wobble which is completely determined by the optical variability data. In this
paper we use the optical data to determine the basic properties of the
helical jet, its opening angle as well as the direction at a given time. Then
there is only the wavelength of the helix and the flow speed to be determined.
These are found by fitting the model to the mm wave and the cm wave jet
orientation data.

Note that in this model there are two flow speeds, the jet flow with
$\Gamma=14$, and a slower flow with $\Gamma\sim3.5$ that carries the
information about the jet wobble from the base of the jet outward. 
\rv{This requires a spine-sheath structure for the jet
\citep{har2007,har2007b},
with the sheath bulk flow moving with $\Gamma\sim3.5$. However, we cannot
rule out the possibility that the information of the jet wobble is carried
by e.g. an oscillating disturbance in the sheath. In this case the wave motion
may propagate with $\Gamma\sim3.5$ with actual bulk flow of the sheath 
being much slower. The latter picture is qualitatively supported by considering
3D-animations of the entire jet given by the model. 
These demonstrate the visibly oscillatory character of the resulting 3D-jet.
}
\rvd{
Thus we require a spine-sheath structure for the jet
\mbox{\citep{har2007,har2007b}}. 
}
The fast jet is narrow; its value
is not specified in the model, but observationally its half opening angle
should be $\theta_\mathrm{fast}=0.8\pm0.4$ degrees 
\citep{jor2005}. The slow jet
is wide, with the half-opening angle of $\theta_\mathrm{slow}=4$ degrees.
These opening angles follow the mean relation between $\theta$ and
$\Gamma$ found by \cite{pus2009}.

It is interesting that this simple model explains the apparently complex dual
jet seen in observations by 
\citet{tat2013}.

\section{Summary}

The complex behavior of the radio jet has been shown to result from a model
with three main assumptions. (1) OJ287 is a binary black hole system whose
orbit is described by the celestial mechanical solution of parameters. This
solution is complemented by the missing parameter, the inclination of the
binary with respect to the disk of the primary, determined from recent
spectroscopy. (2) The jet propagates along the rotation axis of the inner disk.
(3) The jet has spine-sheath structure, and the spine follows a helical path.

The model allows us to determine the jet parameters. There is prior information
on some, but not all of the parameters. Thus the model increases our knowledge
of the jet in this case. The value for the spine $\Gamma=14$ agrees with
previous determinations, as does the typical sheath viewing angle 2 degrees.
The sheath opening angle of 4 degrees was not previously known, neither the
Lorentz $\Gamma=3.5$ for the sheath. The helical wavelength of \rv{$\sim70$ pc} 
is also
determined for the first time, even though it could also be determined from the
high resolution maps of \citet{tat2013}.
It is also interesting that the period of the helical motion agrees in order of
magnitude with the Kozai cycle of inner accretion disk. Moreover, the fact that
the variation of optical polarization follows the model allows us to state
definitely that the magnetic field is parallel to the jet in the optical
emission region, a conclusion which was only tentative in earlier work 
\citep{val2012b}.

\begin{acknowledgements}
P.~Pihajoki is supported by the Magnus Ehrnrooth foundation.
\rv{We thank the anonymous referee for helpful comments improving the article.}
\end{acknowledgements}

\bibliographystyle{aa}
\bibliography{mn-jour,references}

\end{document}